# Dataset after seven years simulating hybrid energy systems with Homer Legacy

Alexandre Beluco [1], Frederico A. During Fº [1], Lúcia M. R. Silva [1],
Jones S. Silva [1], Luís E. Teixeira [1], Gabriel Vasco [1], Fausto A. Canales [2],
Elton G. Rossini [3], José de Souza [4], Giuliano C. Daronco [5] and Alfonso Risso [1]

1 Instituto de Pesquisas Hidráulicas (IPH)as, Universidade Federal do Rio Grande do Sul (UFRGS), Porto Alegre, Rio Grande do Sul (RS), Brazil
2 Department of Civil and Environmental, Universidad de la Costa, Baranquilla, Atlantico, Colombia
3 Universidade Estadual do Rio Grande do Sul (UERGS), Porto Alegre, RS, Brazil
4 Fundação Liberato Salzano Vieira da Cunha, Novo Hamburgo, RS, Brazil
5 Companhia Riograndense de Saneamento (CORSAN), Santa Rosa, RS, Brazil

Corresponding author: albeluco@iph.ufrgs.br

**Abstract.** Homer Legacy software is a well-known software for simulation of small hybrid systems that can be used for both design and research. This dataset is a set of files generated by Homer Legacy bringing the simulation results of hybrid energy systems over the last seven years, as a consequence of the research work led by Dr. Alexandre Beluco, Federal University of Rio Grande do Sul, in southern Brazil. The data correspond to thirty papers published in the last seven years. Two of them describe hydro PV hybrid systems with photovoltaic panels operating on the water surface of reservoirs. One of these twelve papers suggests the modeling of hydropower plants with reservoirs and the other the modeling of pumped hydro storage, and a third still uses these models in a place that could receive both the two types of hydroelectric power plant. The other simulated hybrid systems include wind turbines, diesel generators, batteries, among other components. This data article describes the files that integrate this dataset and the papers that have been published presenting the hybrid systems under study and discussing the results. The files that make up this dataset are available[*] on Mendeley Data repository at *dx.doi.org/10.17632/ybxsttf2by.2*.

## Introduction

Hybrid energy systems can take on very complex behavior, considering the characteristics presented by the energy availability of renewable resources, by the energy demand profiles of the consumers, by the energy storage devices and by the possible combinations between exploited renewable energy resources and between devices for energy storage.

The simulation of different hybrid systems with different components, operating in different locations and conditions, complemented by the observations (whenever possible) of the receptivity of these systems by their users, contributes to a better understanding of their characteristics and to better conceptions of the systems to be designed in the future.

In this context, Homer Legacy (HomerEnergy, 2007) is a well-known software (Connolly et al., 2010) and is a powerful tool for simulating hybrid energy systems. Homer Legacy (Lambert, Gilman & Lilienthal, 2005; Lilienthal, Lambert & Gilman, 2004) simulates all combina-tions of optimization variables to identify which combination leads to the lowest total net present cost over the period considered for project analysis, usually 20-25 years, and repeats these simulations for all sensitivity variables to perform a sensitivity analysis.

---

[*] Available during submission at data.mendeley.com/datasets/ybxsttf2by/draft?a=845e5e7a-bcef-4144-ba0f-4ca124ff71fa.

Homer software was developed in the early years of this century by members of the National Renewable Energy Laboratory (NREL), until version 2.68. The team directly responsible for Homer then formed HomerEnergy to better support users and to develop new versions of the software, meeting a growing demand for services beyond what could be handled by a government lab. The latest version even distributed by NREL has become the Legacy version, still distributed free for educational and academic purposes.

This paper presents files obtained with the Homer Legacy software for nine different case studies and for three methods for simulation of specific hybrid systems. One such method (Canales & Beluco, 2014) describes how to simulate hydro power plants with reservoir and the other (Canales, Beluco & Mendes, 2017) describes how to simulate pumped storage plants, since Homer simulates only run-of-the-river hydropower.

The third (Silva & Beluco, 2018) of these methods for the use of Homer Legacy presents a viability space for new technologies for generating energy from renewable resources. This viability space, as a 'viability window', is a target to be pursued in order to make technical and economically viable new technologies. This paper shows the application to the ocean wave energy case on the southern coast of Brazil.

The two methods described above were applied in a location where it would be possible to build both a conventional hydroelectric power plant with water reservation and a pumped storage hydropower plant. The comparison (Canales, Beluco & Mendes, 2015) indicated, for similar capacities and performances, lower environmental impacts associated with the reversible hydropower plant.

Among the other case studies, the first three of them (Silva, Cardoso & Beluco, 2012; Beluco & Ponticelli, 2014; Beluco et al., 2013) deal with the insertion of PV modules and-or other components or new fuels into previously existing diesel hybrid systems and the fourth (Benevit et al., 2016) of them studies the subtle influence of different wind energy availability profiles on the performance of a wind diesel hybrid system.

The article by During F$^o$. et al. (2018) study the influence of temporal complementarity in time [according to the concept of energetic complementarity discussed by Jurasz et al. (2019)] on the energy storage in batteries in PV hydro hybrid systems, applying a specific method (Beluco, Souza & Krenzinger, 2012) appropriate for the analysis of hybrid systems based on complementary energy resources.

The paper by Teixeira et al. (2015) evaluates the feasibility of a PV hydro hybrid system operating in a dam for water supply in southern Brazil. The hydropower plant is feasible only with the use of pumps as turbines and the PV modules must be installed on floating structures on the surface of the reservoir. This feasibility evaluation is similar to the one that was undertaken for the pumped storage plant proposed by Risso et al. (2017) in their study.

Finally, two papers consider an alternative to a hydropower plant whose construction was started in the 1970s and then interrupted after the completion of the dam. The first (Vasco et al.. 2019a) proposes an engine room at the base of the dam, unlike the original proposal, with PV modules on floating structures on the surface of the reservoir. The other paper (Vasco et al., 2019b) evaluates the use of the reservoir's small power storage capacity to obtain a hybrid system with better performance.

This article consists of four sections. This introduction briefly presents the topics covered in the papers that were published from the Homer files composing this dataset. The next section describes the thirty Homer files contained in this dataset. The subsequent section briefly presents the method used to build the Homer files composing this data set and the fourth and final section outlines some basic information on how to get the Homer software, Legacy version, and how to use it and how to access the files in this dataset.

## Data description

The files composing this dataset are available online in the Mendeley Data repository under a specific DOI (Beluco et al., 2020), with each file also identified with a specific DOI. Table 1, below, presents some specifications of this dataset, and subsequently each of the files is briefly described. The next section describes how the data in these files was obtained.

**Table 1.** Specifications of the dataset.

| Feature | Description |
| --- | --- |
| Subject | Engineering |
| Specific subject area | Renewable Energy |
| Type of data | Results of simulations performed by the software Homer Legacy, in the format of *hmr* files (compatible with this specific software) |
| Description of data set | 30 *hmr* files corresponding to 12 cases studied, including proposed methods and case studies, with results presented in 12 papers. |
| Data format | Raw data, automatically analyzed by Homer Legacy when opened |
| Input data | Energy availability data of the resources used in each case, consumer demand profiles, technical specifications of the components of the hybrid systems |
| Output data | One-year operation simulations of hybrid systems for all combinations of optimization and sensitivity variables, and accounting for all costs for the period of analysis |
| Data source location | Simulation data were obtained for operational or design hybrid energy systems at some locations along the territory of the State of Rio Grande do Sul, the southernmost State of Brazil. |
| Data accessibility | Data available in Mendeley Data repository: 10.17632/ybxsttf2by.2. During submission, available at https://data.mendeley.com/datasets/ybxsttf2by/draft?a=845e5e7a-bcef-4144-ba0f-4ca124ff71fa. |
| Related research article | Silva et al. 2012. Int J Photoenergy, v.2012, #384153.<br>Beluco et al. 2013. Comp W Energy Envrn Eng, v.2, n.2, p.43-53.<br>Canales and Beluco. 2014. J Ren Sust Energy, v.6, #043131<br>Beluco and Ponticelli. 2014. Int J Ren En Tech, v.5, n.3, p.229-250.<br>Canales et al. 2015. J En Stor, v.4, p.96-105.<br>Teixeira et al. 2015. J Pow En Eng, v.3, n.9, p.70-83.<br>Benevit et al. 2016. J Pow En Eng, v.4, n.8, p.38-48.<br>Canales et al. 2017. Int J Sust En, v.36, n.7, p.654-667.<br>Silva and Beluco 2018. Curr Alt En, v.2, n.2, p.112-122.<br>During Fo. et al. 2018. Comp W En Envrn Eng, v.7, n.3, p.142-159.<br>Vasco et al. 2019. Comp W En Envrn Eng, v.8, n.2, p.41-56.<br>Vasco et al. 2019. Sm Grid Ren En, v.10, n.4, p.83-97. |

The thirty *hmr* files constituting this dataset, corresponding to twelve papers, are organized in chronological order, as briefly described below:

- *#01-silva-et-al-2012.hmr* - a PV wind diesel hybrid system with energy storage in batteries and water and environment heating, 288 optimization and 5880 sensitivity values, results presented by Silva, Cardoso & Beluco (2012);
- *#02-beluco-et-al-2013-A.hmr* - a PV hydro diesel hybrid system connected to the grid, 48 optimization values and 1536 sensitivity values, results presented by Beluco et al. (2013);
- *#02-beluco-et-al-2013-B.hmr* - a PV hydro diesel hybrid system connected to the grid, 48 optimization values and 1536 sensitivity values, results presented by Beluco et al. (2013);
- *#03-canales-beluco-2014-1.hmr* - a wind hydro diesel hybrid system with pumped storage capacity, 15 optimization values and 3 sensitivity values, results presented by Canales & Beluco (2014);

- *#03-canales-beluco-2014-2.hmr* - a wind hydro diesel hybrid system with pumped storage capacity, 24 optimization values and 3 sensitivity values, results presented by Canales & Beluco (2014);
- *#04-beluco-ponticelli-2014-fig1.hmr* - a wind diesel hybrid system, 15 optimization values and 1872 sensitivity values, results presented by Beluco & Ponticelli (2014);
- *#04-beluco- ponticelli-2014-fig6a-b100.hmr* - a PV wind biodiesel hybrid system with energy storage in batteries, 1750 optimization values and 1000 sensitivity values, results presented by Beluco & Ponticelli (2014);
- *#04-beluco- ponticelli-2014-fig6a-dsl.hmr* - a PV wind diesel hybrid system with energy storage in batteries, 1750 optimization values and 1000 sensitivity values, results presented by Beluco & Ponticelli (2014);
- *#04-beluco- ponticelli-2014-fig6b.hmr* - a PV wind diesel biodiesel hybrid system with energy storage in batteries, 3500 optimization values and 50 sensitivity values, results presented by Beluco & Ponticelli (2014);
- *#05-canales-et-al-2015-Sys3PH.hmr* - a wind hydro hybrid system with pumped storage capacity, 12474 optimization values and 96 sensitivity values, results presented by Canales, Beluco & Mendes (2015);
- *#05-canales-et-al-2015-Sys3Res.hmr* - a wind hydro hybrid system with energy storage capacity in the water reservoir, 1496 optimization values and 48 sensitivity values, results presented by Canales, Beluco & Mendes (2015);
- *#06-teixeira-et-al-2015-2000.hmr* - a PV hydro hybrid system designed for operation at a dam for water supply, with the capital cost of the PV modules at US$ 2000 per kW, 300 optimization values and 1440 sensitivity values, results presented by Teixeira et al. (2015);
- *#06-teixeira-et-al-2015-2500.hmr* - a PV hydro hybrid system designed for operation at a dam for water supply, with the capital cost of the PV modules at US$ 2500 per kW, 300 optimization values and 1440 sensitivity values, results presented by Teixeira et al. (2015);
- *#06-teixeira-et-al-2015-3000.hmr* - a PV hydro hybrid system designed for operation at a dam for water supply, with the capital cost of the PV modules at US$ 3000 per kW, 300 optimization values and 1440 sensitivity values, results presented by Teixeira et al. (2015);
- *#06-teixeira-et-al-2015-3500.hmr* - a PV hydro hybrid system designed for operation at a dam for water supply, with the capital cost of the PV modules at US$ 3500 per kW, 300 optimization values and 1440 sensitivity values, results presented by Teixeira et al. (2015);
- *#06-teixeira-et-al-2015-4000.hmr* - a PV hydro hybrid system designed for operation at a dam for water supply, with the capital cost of the PV modules at US$ 4000 per kW, 300 optimization values and 1440 sensitivity values, results presented by Teixeira et al. (2015);
- *#07-benevit-et-al-2016-180.hmr* - a wind diesel hybrid system with energy storage in batteries, simulated with Weibull shape parameter equal to 1.80, 360 optimization values and 108 sensitivity values, results presented by Benevit et al. (2016);
- *#07-benevit-et-al-2016-210.hmr* - a wind diesel hybrid system with energy storage in batteries, simulated with Weibull shape parameter equal to 2.10, 360 optimization values and 108 sensitivity values, results presented by Benevit et al. (2016);
- *#07-benevit-et-al-2016-240.hmr* - a wind diesel hybrid system with energy storage in batteries, simulated with Weibull shape parameter equal to 2.40, 360 optimization values and 108 sensitivity values, results presented by Benevit et al. (2016);
- *#07-benevit-et-al-2016-270.hmr* - a wind diesel hybrid system with energy storage in batteries, simulated with Weibull shape parameter equal to 2.70, 360 optimization values and 108 sensitivity values, results presented by Benevit et al. (2016);
- *#07-benevit-et-al-2016-300.hmr* - a wind diesel hybrid system with energy storage in batteries, simulated with Weibull shape parameter equal to 3.00, 360 optimization values and 108 sensitivity values, results presented by Benevit et al. (2016);
- *#08-canales-et-al-2017.hmr* - a wind hydro hybrid system with energy storage in the reservoir of the hydropower plant, 224 optimization values and 72 sensitivity values, results presented by Canales, Beluco & Mendes (2017);
- *#09-silva-beluco-2018.hmr* - a PV wind diesel hybrid system connected to the grid, includ-

ing an ocean wave power plant, 512 optimization values and 450 sensitivity values, results presented by Silva & Beluco (2018);
- *#10-during-et-al-2018-d000.hmr* - a PV hydro hybrid system with energy storage capacity in batteries, with time-complementarity index equal to 0.00, 1476 optimization values and 87 sensitivity values, results presented by During F⁰. et al. (2018);
- *#10-during-et-al-2018-d090.hmr* - a PV hydro hybrid system with energy storage capacity in batteries, with time-complementarity index equal to 0.50, 1476 optimization values and 87 sensitivity values, results presented by During F⁰. et al. (2018);
- *#10-during-et-al-2018-d120.hmr* - a PV hydro hybrid system with energy storage capacity in batteries, with time-complementarity index equal to 0.67, 1476 optimization values and 87 sensitivity values, results presented by During F⁰. et al. (2018);
- *#10-during-et-al-2018-d150.hmr* - a PV hydro hybrid system with energy storage capacity in batteries, with time-complementarity index equal to 0.83, 1476 optimization values and 87 sensitivity values, results presented by During F⁰. et al. (2018);
- *#10-during-et-al-2018-d180.hmr* - a PV hydro hybrid system with energy storage capacity in batteries, with time-complementarity index equal to 1.00, 1476 optimization values and 87 sensitivity values, results presented by During F⁰. et al. (2018);
- *#11-vasco-et-al-2019.hmr* - a PV hydro hybrid system connected to the grid, designed for an unfinished hydro power plant, 486 optimization and 567 sensitivity values, results presented by Vasco et al. (2019a);
- *#12-vasco-et-al-2019.hmr* - a PV hydro hybrid system connected to the grid, with energy storage in the water reservoir, 18 optimization values and 72 sensitivity values, results presented by Vasco et al. (2019b).

**Methods**

The Homer Legacy was used to generate the data available in this dataset and this software is available for free at the link provided by HomerEnergy (2007), as described in the next section. For each of the hybrid energy systems that were simulated and whose results appear in the files described above, the software received energy availability data for each of the energy resources employed, the software also received the demand profile of the consumer loads and also received technical information on each of the components of the hybrid systems that were simulated. The Homer software then performs simulations of these hybrid systems for all values of the optimization variables and the sensitivity variables, as explained by Lambert, Gilman & Lilienthal (2005) and Lilienthal. Lambert & Gilman (2004). The files indicated above then present the results of all these simulations.

**Software Homer, version Legacy - how to get it and how to use it**

Homer allows the selection of components that form the hybrid energy system to be studied, simulates this hybrid system for a full year, at 8760 hours of a year, simulates this system for all combinations of optimization variables choosing as optimal solution that solution presenting the lowest total net present cost, and repeats these simulations for all combinations of sensitivity variables, leading to an optimization space. The simulations are carried out for one year, but the economic feasibility takes into account the time established for the evaluation, usually 20-25 years. Homer can be well understood by consulting Lambert, Gilman & Lilienthal (2005) and Lilienthal. Lambert & Gilman (2004).

The version Legacy of software Homer corresponds to the latest version provided by NREL [National Renewable Energy Laboratory (U.S. Department of Energy, 2019)]. The team responsible for Homer's development formed HomerEnergy and went on to develop more advanced versions in a commercial scheme. Legacy version is still available free for educational and academic activities after registration on the HomerEnergy website (HomerEnergy, 2019) and the completion of a specific form. Homer was part of the software package made available by NREL until 2007 and later distributed by HomerEnergy (still free of charge).

The files in this data set can be accessed directly with Homer and the results can be explored by searching for details that are not discussed in the papers cited above. The Getting

Started Guide (Lilienthal. Lambert & Gilman, 2011) provides complete instructions for general operation of Homer and for detailed exploration of results, including fine-tuning a project. As commented in the text of the first section of the Getting Started Guide, "it should take about an hour to complete this exercise". The Getting Started Guide can also be accessed in the third line of the Help item in the main menu of the Homer software.

**Acknowledgments**
This work was developed as a part of the activities of the Research Group on Renewable Energy and Sustainability of the Instituto de Pesquisas Hidráulicas, Universidade Federal do Rio Grande do Sul, in southern Brazil. The authors acknowledge the support received by the institution. The research work of the first author is supported by CNPq (proc. n. 312941/2017-0).

**Competing interests**
The authors have no competing interests to declare.